\documentclass[12pt]{iopart}
\usepackage{graphicx}

\begin{document}

\title{Analytical evidence for the absence of spin glass transition on self-dual lattices}
\author{Masayuki Ohzeki and Hidetoshi Nishimori}
\address{Department of Physics, Tokyo Institute of Technology, Oh-okayama,
Meguro-ku, Tokyo 152-8551, Japan} 
\begin{abstract}
We show strong evidence for the absence of a finite-temperature spin glass transition for the random-bond Ising model on self-dual lattices.
The analysis is performed by an application of duality relations, which enables us to derive a precise but approximate location of the multicritical point on the Nishimori line.
This method can be systematically improved to presumably give the exact result asymptotically.
The duality analysis, in conjunction with the relationship between the multicritical point and the spin glass transition point for the symmetric distribution function of randomness, leads to the conclusion of the absence of a finite-temperature spin glass transition for the case of symmetric distribution.
The result is applicable to the random bond Ising model with $\pm J$ or Gaussian distribution and the Potts gauge glass on the square, triangular and hexagonal lattices as well as the random three-body Ising model on the triangular and the Union-Jack lattices and the four dimensional random plaquette gauge model.
This conclusion is exact provided that the replica method is valid and the asymptotic limit of the duality analysis yields the exact location of the multicritical point.
\end{abstract}
\ead{mohzeki@stat.phys.titech.ac.jp}
\submitto{J. Phys. Math. Theor.}
\maketitle
\normalsize

\section{\protect\normalsize Introduction}
Properties of finite-dimensional spin glasses are still under active current investigations after thirty years since mean field analyses of the basic model \cite{EA,SK}. 
The difficult problem of whether or not the mean-field predictions apply to realistic finite-dimensional systems is still largely unsolved. 
One of the outstanding problems is the existence or absence of the spin glass phase. 
Most of the current investigations on this problem are carried out using numerical methods \cite{Rev1,Rev2,Rev3}.

Two typical spin glass models have been examined extensively, the Gaussian and $\pm J$ Ising models. 
In three dimensions, researchers have arrived at the consensus that there are finite-temperature spin glass transitions in both models. 
On the other hand, in two dimensions, numerical investigations show evidence that there would be no finite-temperature spin glass transition in both models \cite{Gauss1,Gauss2,Gauss3,GaussPMJ1,GaussPMJ2,Gauss4,PMJ1,PMJ2,PMJ3}.
Unfortunately no reliable analytical evidence for the problem of the existence or absence of the spin glass phase in finite dimensions has been established.

Very little systematic analytical work for finite-dimensional spin glasses exists.
An exception is a technique based on the gauge symmetry to derive the exact value of the internal energy, a rigorous upper bound of the specific heat, several set of rigorous inequalities and exact relations in a special subspace, known as the Nishimori line \cite{HN81,HNbook}. 
In the present study, we develop an argument by the gauge symmetry in conjunction with the duality and the replica method to study the problem whether or not a finite-temperature spin glass transition exists in two dimensions. 
We analyze the problem of the spin glass transition point for the $\pm J$ Ising model and the Gaussian Ising model on self-dual lattices by means of the duality \cite{KW,WuWang}.
The theory is applicable directly to the square lattice, but the triangular and hexagonal lattices can also be reduced to be self-dual using the duality in conjunction with the star-triangle transformation \cite{WuPotts}.
In the present study, we arrive at the conclusion that no finite-temperature spin glass transition exits in the symmetric distribution of randomness, for example, $p=1/2$ for the $\pm J$ Ising model.
The result is justified under the validity of the replica method.
It should also be remembered that the prediction of the duality method for the transition point is expected to be exact only in the asymptotic limit of large cluster size in the sense we shall define in the following sections.

This paper is organized as follows. 
In the next section, we recall the basic formulations of the duality in spin glasses with the replica method and the relation derived by the gauge symmetry to set a stage to show the absence of a spin glass phase in the following section. 
We develop an analytical argument to derive our result in section 3.
In this section, we show the limit of applicability of the present result.
The final section is devoted to conclusion and discussions. 

\section{Duality and gauge symmetry}
\begin{figure}
\begin{center}
\includegraphics[width=100mm]{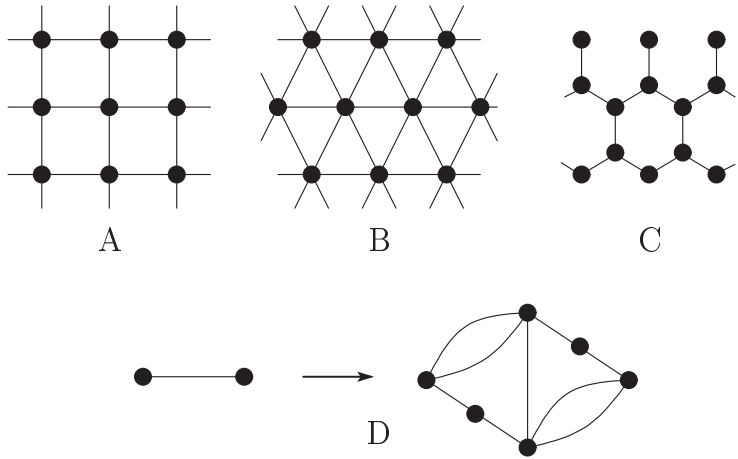}
\end{center}
\caption{{\small Self-dual lattices. (A) Square lattice. (B) Triangular lattice. (C) Hexagonal lattice. (D) Hierarchcial lattice.
The triangular and hexagonal lattices become self-dual by means of the duality combined with the star-triangle transformation \cite{WuPotts}.
Construction of a hierarchical lattice starts from a single bond, and we iterate the process to substitute the single bond with the unit cell of a complex structure \cite{HL1,HL2,HL3}.
}}\label{fig1}
\end{figure}

Let us review several known facts in the present section to fix the notation and prepare for the developments in the next section.
The following arguments are applicable to several spin glass models on self-dual lattices as shown in figure \ref{fig1}.
We take the $\pm J$ Ising model on the square lattice as an example here for simplicity.
It is straightforward to generalize the following arguments to other spin glass models as will be explained in the next section.
\subsection{$\pm J$ Ising model}
The Hamiltonian of the $\pm J$ Ising model is given by
\begin{equation}
H=-\sum_{\langle ij\rangle }J_{ij}S_{i}S_{j},
\end{equation}%
where $S_{i}$ is the Ising spin taking $\pm 1$, and $J_{ij}(=\pm J)$ denotes the quenched random coupling.
The summation is taken over nearest neighbouring pairs of sites. 
The partition function with fixed randomness is
\begin{equation}
Z(K,\{\tau_{ij} \})=\sum_{\{S_{i}\}}\prod_{\langle ij\rangle }\mathrm{e}^{K\tau_{ij}S_{i}S_{j}},
\end{equation}
where $K=\beta J$ is the coupling constant, and $\tau _{ij}(=\pm 1)$ is the sign of the random coupling $J_{ij}$. 
The distribution function of $\tau_{ij}$ is
\begin{equation}
P(\tau _{ij})=p\delta (\tau _{ij}-1)+(1-p)\delta (\tau _{ij}+1)=\frac{\mathrm{e}^{K_{p}\tau _{ij}}}{2\cosh K_{p}},
\end{equation}%
where we defined $K_p$ by $\exp \left( -2K_{p}\right) =(1-p)/p$.
\subsection{Duality and multicritical point of spin glasses}
We here review the analysis of the location of the multicritical points by the duality in spin glasses \cite{NN,MNN}.
The multicritical point is connected with a critical point for $p=1/2$ $(K_p=0)$ by a relation introduced in section 2.4.

We apply the replica method to the $\pm J$ Ising model.
The $n$-replicated partition function after the configurational average is
\begin{equation}
Z_n(K,K_p) = \frac{1}{(2\cosh K_p)^{N_B}}\sum_{\{\tau_{ij}\}}\prod_{\langle
ij \rangle}\mathrm{e}^{K_p\tau_{ij}}Z(K,\{\tau_{ij}\})^n,  \label{OP}
\end{equation}
where $n$ stands for the number of replicas.
We generalize the duality argument to this $n$-replicated $\pm J$ Ising model \cite{NN,MNN}.
For this purpose it is useful to define the edge Boltzmann factor $x_k ~(k = 0, 1, \cdots, n)$, which represents the configuration-averaged Boltzmann factor for interacting spins with $k$ antiparallel spin pairs among $n$ nearest-neighbour pairs for a bond (edge).
The duality gives the relationship of the partition functions with different values of the edge Boltzmann factor as given by 
\begin{equation}
Z_n(x_0,x_1,\cdots,x_n) = Z_n(x^*_0,x^*_1,\cdots,x^*_n).  \label{PF2}
\end{equation}
The dual edge Boltzmann factors $x_k^*$ are defined by the discrete multiple Fourier transforms of the original edge Boltzmann factors, which are simple combinations of plus and minus of the original Boltzmann factors in the case of Ising spins.
Two principal Boltzmann factors $x_0$ and $x^*_0$ are the most important elements of the theory \cite{NN,MNN}:
\begin{eqnarray}
x_0(K,K_p) &=& \frac{\cosh \left(K_p + n K \right)}{\cosh K_p }, \\
x^*_0(K,K_p) &=& \left( \sqrt{2} \cosh K \right)^n.
\end{eqnarray}
We extract these principal Boltzmann factors from the partition functions in equation (\ref{PF2}) to measure the energy from the all-parallel spin configuration.
The result is written as, using the normalized edge Boltzmann factors $u_j = x_j/x_0$ and $u^*_j = x^*_j/x^*_0$,
\begin{equation}
{x_0(K,K_p)}^{N_B}z_n(u_1,u_2,\cdots,u_n) = {x^*_0(K,K_p)}%
^{N_B}z_n(u^*_1,u^*_2,\cdots,u^*_n), \label{PF1}
\end{equation}
where $z_n(u_1,\cdots)$ and $z_n(u^*_1,\cdots)$ are defined as $Z_n/x^{N_B}_0$ and $Z_n/(x^{*}_0)^{N_B}$.

The duality identifies the critical point under the assumption of a unique phase transition. 
The critical point is given as the fixed point of the duality transformation and is known to yield the exact critical point for a simple ferromagnetic system on the square lattice \cite{KW,WuWang}.
In order to obtain the multicritical point of the present replicated spin glass system, we set $K=K_p$, which defines the Nishimori line (NL) on which the multicritical point is expected to lie.
Since $z_n$ is a multivariable function, there is no fixed point of the duality relation in the strict sense which satisfies $n$ conditions simultaneously, $u_1(K)=u^*_1(K),u_2(K)=u^*_2(K),\cdots,u_n(K)=u^*_n(K)$.
This is in sharp contrast to the non-random Ising model, in which the duality is a relation between single-variable functions.
We nevertheless set a hypothesis that a single equation $x_0(K,K)=x_0^*(K,K)$ gives the location of the multicritical point for any replica number $n$ \cite{NN,MNN,TN,TSN,NO,NStat}.
The quenched limit $n \to 0$ for the equation $x_0(K,K)=x_0^*(K,K)$ then yields
\begin{equation}
-p \log p - (1-p) \log (1-p) = \frac{1}{2}\log 2.
\end{equation}
The solution to this equation is $p_c = 0.889972 (\approx 0.8900)$.

This value of $p_c$ is very close to numerical results but shows small deviations on several self-dual hierarchical lattices, for which numerically exact results can be derived, as in table \ref{Con} \cite{ONB}.
Such discrepancies may come from the following two facts.
One is that the condition $x_0(K,K)=x_0^*(K,K)$ is different from the strict fixed-point condition and the other is that we have considered only the quantity defined on a single bond, the principal Boltzmann factors $x_0$ and $x_0^{*}$, which does not necessarily reflect the effects of frustration inherent in spin glasses.
The improvements with these points taken into account will be explained in the next section, following \cite{ONB,MCP}.
\subsection{Improved method}
\begin{figure}
\begin{center}
\includegraphics[width=130mm]{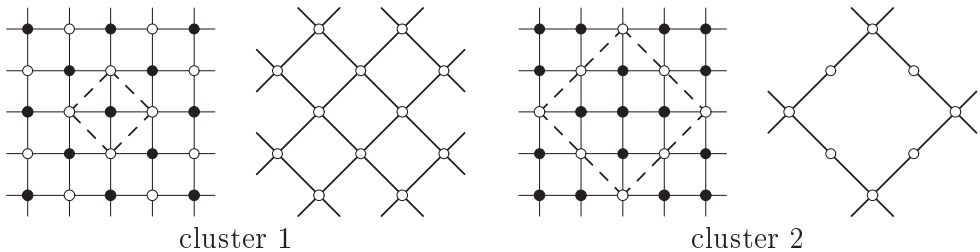}
\end{center}
\caption{A cluster is the unit plaquette encircled by white spins.
The spins marked black on the original lattice are traced out to yield interactions among white spins in clusters.}\label{fig2}
\end{figure}

As shown in figure \ref{fig2}, let us consider to sum over a part of the spins, called a cluster below, on the square lattice to deal with the effects of frustration rather than a single bond considered in the naive approach described in the previous section.
Then, the set of clusters must be chosen to cover the whole lattice under consideration as in figure \ref{fig2}, where two examples of the choice of clusters are depicted. 
The partition function can now be expressed as a function of the configuration of spins around a cluster (marked white in figure \ref{fig2}). 
The duality is written as
\begin{equation}
Z_{n}^{(s)}(x_{0}^{(s)},x_{1}^{(s)},\cdots )=Z_{n}^{(s)}(x_{0}^{\ast (s)},x_{1}^{\ast (s)},\cdots ).
\end{equation}%
The superscript $s$ stands for the type of the cluster that one chooses.
The quantity $x_{k}^{(s)}$ is the local Boltzmann factor including many-body interactions generated by the summation over spins marked black in figure \ref{fig2}.
We define the principal Boltzmann factors $x_{0}^{(s)}$ and its dual $x_{0}^{\ast (s)}$ as those with all spins surrounding the cluster in the up state.
We assume that a single equation gives the accurate location of the multicritical point for any number of $n$, similarly to the naive conjecture, 
\begin{equation}
x_{0}^{(s)}(K,K)=x_{0}^{\ast (s)}(K,K).  \label{Imporved}
\end{equation}%
This is the improved method to predict the location of the multicritical point with higher precision than the naive conjecture.

\begin{table}[htbp]
\begin{center}
\begin{tabular}{ccc}
\hline
$p_c$ (conjecture) & $p_c$ (improved) & $p_c$ (numerical) \\
\hline
$0.8900$ & $0.8920$ & $0.8915(6)$ \\
$0.8900$ & $0.8903$ & $0.8903(2)$ \\
$0.8900$ & $0.8892$ & $0.8892(6)$ \\
$0.8900$ & $0.8895$ & $0.8895(6)$ \\
$0.8900$ & $0.8891$ & $0.8890(6)$ \\
\hline
\end{tabular}
\end{center}
\caption{Comparison of the naive conjecture, improved method and numerical estimations for self-dual hierarchical lattices \cite{ONB}.}
\label{Con}
\end{table}
The improved method for the $\pm J$ Ising model indeed has given the results in excellent agreement with the exact estimations within numerical error bars on several self-dual hierarchical lattices as summarized in table \ref{Con} \cite{ONB,MCP}.
In addition, recent numerical investigations on the square lattice have given $p_c = 0.89081(7)$ \cite{Hasen} and $p_c = 0.89061(6)$ \cite{Queiroz}, while the improved method has estimated $p_c = 0.890725$ by cluster 1 of figure $\ref{fig2}$, and $p_c = 0.890822$ by cluster 2 \cite{MCP}.
If we deal with clusters of larger sizes, the improved method should show systematic improvements toward the exact answer on the location of the multicritical point.
We use this property of the improved method to show that the spin-glass transition temperature would be zero on self-dual lattices, which means the absence of a finite-temperature spin glass transition.
\subsection{Relation between different replica numbers}
Another important piece of information is the relation that the $(n+1)$-replicated system with $p = 1/2$ ($K_p = 0$) is equivalent to the $n$-replicated system on the NL \cite{Georges}. 

After the gauge transformation and summation over gauge variables, the exponential in equation (\ref{OP}) turns to a partition function with coupling $K_{p}$ \cite{HN81,HNbook}: 
\begin{equation}
Z_{n}(K,K_{p})=\frac{1}{2^{N_{s}}(2\cosh K_{p})^{N_{B}}}\sum_{\{\tau_{ij}\}}Z(K_{p},\{\tau _{ij}\})\cdot Z(K,\{\tau _{ij}\})^{n},
\end{equation}%
where $N_{s}$ is the total number of spins. It readily follows from this equation that $Z_{n}(K,K)$ and $Z_{n+1}(K,0)$ are essentially equal to each other: 
\begin{equation}
2^{N_{s}}(2\cosh K)^{N_{B}}Z_{n}(K,K)=2^{N_{s}+N_{B}}Z_{n+1}(K,0).
\label{GR}
\end{equation}
This result is quite general since we did not use the properties of a specific lattice. 
If there is a singularity in the partition function $Z_n(K,K)$ on the left-hand side at some point $(K_{c},K_{c})$ for replica number $n$, $Z_{n+1}(K,0)$ on the right-hand side has also a singularity at $(K_{c},0)$ for $n+1$.
According to this relation, it is sufficient to evaluate the location of the multicritical point in the limit $n\rightarrow -1$ in order to study whether or not there is a finite-temperature spin glass transition for $K_p=0$ in the quenched disorder system ($n\rightarrow 0$).
In the following section, we will derive clear evidence for the absence of a finite-temperature spin glass transition for the case of $p=1/2$ $(K_{p}=0)$ by estimating the location of the multicritical point in the limit $n\rightarrow -1$ by the improved method.
The present argument is valid as long as the replica method is reliable, in particular in the limit $n \to -1$.
\section{Absence of a finite-temperature transition for $p=1/2$}
In this section we show that $T_{\rm SG}=0$ for $p=1/2$ $(K_p=0)$ for an arbitrary cluster of the improved method introduced in the previous section.
Since the improved method would give the exact result in the limit of infinitely large clusters because we take the full trace over all spins in the system, except for the boundary spins, to give the exact partition function, we expect our conclusion $T_{\rm SG}=0$ to be valid not as an approximation but as the exact conclusion.

According to the improved method, the location of the multicritical point for the $n$-replicated $\pm J$ Ising model is given by equation (\ref{Imporved}).
The explicit expressions of two principal Boltzmann factors are given as
\begin{equation}
x^{(s)}_0(K,K_p) = \frac{1}{(2\cosh K_p)^{N^{\mathrm{cl.}}_B}}
\sum_{\{\tau_{ij}\}}\prod_{\langle ij \rangle}{}^{\prime }\mathrm{e}^{K_p
\tau_{ij}} \left\{ \sum_{\{S_i\}}{}^{\prime }\prod_{\langle ij
\rangle}{}^{\prime }\mathrm{e}^{K \tau_{ij}S_iS_j} \right\}^n,
\end{equation}
and 
\begin{eqnarray}
x^{*(s)}_0(K,K_p) & = & \frac{1}{(2\cosh K_p)^{N^{\mathrm{cl.}}_B}}
\sum_{\{\tau_{ij}\}}\prod_{\langle ij \rangle}{}^{\prime }\mathrm{e}^{K_p
\tau_{ij}}  \nonumber \\
& & \quad \times \left\{ \frac{1}{\sqrt{2}}\sum_{\{S_i\}}{}^{\prime
}\prod_{\langle ij \rangle}{}^{\prime }\left( \mathrm{e}^{K \tau_{ij}} + S_i
S_j\mathrm{e}^{-K \tau_{ij}}\right) \right\}^n \label{xsd} \\ 
&\equiv& \frac{1}{(2\cosh K_p)^{N^{\mathrm{cl.}}_B}}
\sum_{\{\tau_{ij}\}}\prod_{\langle ij \rangle}{}^{\prime }\mathrm{e}^{K_p
\tau_{ij}} Z^{*(s)}(K,\{\tau_{ij}\})^n,
\end{eqnarray}
where the prime on the summation denotes the condition that all the spins at the perimeter of the cluster are up and the prime on the product represents that the pairs are restricted to those within the cluster. The quantity $N_B^{\mathrm{cl.}}$ is the number of bonds in the cluster.

The principal Boltzmann factor $x^{(s)}_0$ is regarded as the $n$-replicated partition function of the finite-size cluster after the configurational average, which is cut out of the self-dual lattice under consideration: 
\begin{equation}
x^{(s)}_0(K,K_p) = \frac{1}{(2\cosh K_p)^{N^{\mathrm{cl.}}_B}}
\sum_{\{\tau_{ij}\}}\prod_{\langle ij \rangle}{}^{\prime }\mathrm{e}^{K_p
\tau_{ij}} Z^{(s)}(K,\{\tau_{ij}\})^n.
\end{equation}
Gauge transformation and summation over gauge variables reduce the exponential factor in this expression to a partition function of the cluster with coupling $K_p$ \cite{HN81,HNbook},
\begin{equation}
x^{(s)}_0(K,K_p) = \frac{1}{2^{N^{\mathrm{cl.}}_s}(2\cosh K_p)^{N^{\mathrm{%
cl.}}_B}}\sum_{\{\tau_{ij}\}}Z^{(s)}(K_p,\{\tau_{ij}\}) {Z^{(s)}}(K,\{\tau_{ij}\})^n,
\end{equation}
where $N_s^{\mathrm{cl.}}$ denotes the number of spins inside the cluster under consideration.

\begin{figure}
\begin{center}
\includegraphics[width=110mm]{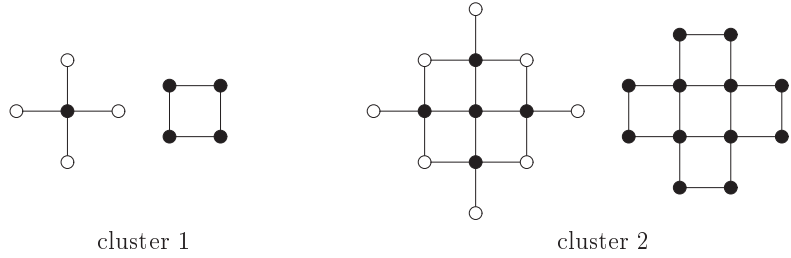}
\end{center}
\caption{{\small Dual pairs for cluster 1 and cluster 2. The same symbols are used as in figure \ref{fig2}.}}\label{fig3}
\end{figure}

The quantity $Z^{*(s)}$ in the dual principal Boltzmann factor $x^{*(s)}_0$, in contrast, is not gauge invariant as seen in equation (\ref{xsd}).
We here apply the duality transformation to $Z^{*(s)}$ in the cluster with all edge spins being up. 
We can then obtain another expression of $Z^{*(s)}$ as \cite{MCP}
\begin{eqnarray}
Z^{*(s)}(K,\{\tau_{ij}\}) &=& 2^{N^{\mathrm{cl.}}_s-N^{\mathrm{cl.}}_B/2-1}
\sum_{\{S_i\}}\prod_{\langle ij \rangle}^{\mathrm{dual}} \mathrm{e}^{K\tau_{ij} S_i S_j} \\
&\equiv& 2^{N^{\mathrm{cl.}}_s-N^{\mathrm{cl.}}_B/2-1} Z_{\rm D}^{(s)}(K,\{\tau_{ij}\}).
\end{eqnarray}
The duality in the cluster allows us to rewrite $Z^{*(s)}(K,\{\tau_{ij}\})$ as the partition function $Z_{\rm D}^{(s)}(K,\{\tau_{ij}\})$ defined on the dual lattice of the cluster under consideration, denoted by the subscript D, as in figure \ref{fig3}.
It is noted that the total number of spins in the dual cluster is given by the number of plaquettes in the original cluster, $N^{\mathrm{cl.}}_p$.
Again the technique of gauge transformation with $N_p^{\rm cl.}$ gauge variables is applicable to rewrite the dual principal Boltzmann factor $x^{*(s)}_0$ as
\begin{equation}
x^{*(s)}_0(K,K_p) = \frac{2^{n(N^{\mathrm{cl.}}_s-N^{\mathrm{cl.}}_B/2-1)}}{%
2^{N^{\mathrm{cl.}}_p}(2\cosh K_p)^{N^{\mathrm{cl.}}_B}} \sum_{\{\tau_{ij}\}}%
{Z_{\rm D}^{(s)}}(K_p,\{\tau_{ij}\}){Z_{\rm D}^{(s)}}(K,\{\tau_{ij}\})^n.
\end{equation}
Our task is thus to solve the following equation derived from $x^{(s)}_0 = x^{*(s)}_0$ with the condition $K=K_p$.
\begin{equation}
\frac{2^{n(N^{\mathrm{cl.}}_s-N^{\mathrm{cl.}}_B/2-1)}}{2^{N^{\mathrm{cl.}%
}_p-N^{\mathrm{cl.}}_s}} \sum_{\{\tau_{ij}\}}{Z_{\rm D}^{(s)}}(K,\{%
\tau_{ij}\})^{n+1} = \sum_{\{\tau_{ij}\}}{Z^{(s)}}(K,\{\tau_{ij}\})^{n+1}.
\label{SG}
\end{equation}
This equation gives the location of the multicritical point, which corresponds to the singularity on the left-hand side of equation (\ref{GR}).
To show the absence of a finite-temperature transition for $K_{p}=0$ in the quenched system ($n\rightarrow 0$), we consider the limit of $n\rightarrow -1$ in the above equation, since the multicritical point for $n\rightarrow -1$ is equivalent to the critical point for $K_{p}=0$ in $n\rightarrow 0$ due to the relation (\ref{GR}).
It should be noted that the two partition functions $Z^{(s)}$ and $Z_{\mathrm{D}}^{(s)}$ do not have any singularity since they are given by the summation over spins of finite-size systems. 

Let us assume that there is a finite-temperature transition $T_{\rm SG}=1/K_{\rm SG}$. 
Then the partition functions $Z^{(s)}$ and $Z_{\mathrm{D}}^{(s)}$ on the original and dual clusters have some finite values. 
Equation (\ref{SG}) then reduces to, in the limit of $n\rightarrow -1$, 
\begin{equation}
2^{N_{p}^{\mathrm{cl.}}-N_{B}^{\mathrm{cl.}}/2-1} = 1.  \label{last}
\end{equation}%
The fact that $N_{p}^{\mathrm{cl.}} \neq N_{B}^{\mathrm{cl.}}/2 +1 $ for any finite-size cluster cut out of the self-dual lattices leads us to the conclusion that equation (\ref{last}) can not be satisfied.
In this sense, the spin glass transition point of the $\pm J$ Ising model on a self-dual lattice goes to zero $T_{\rm SG} \to 0$ as $n \to -1$.
By taking the asymptotic limit of large clusters, we expect the result to be exact.
\begin{figure}
\begin{center}
\includegraphics[width=80mm]{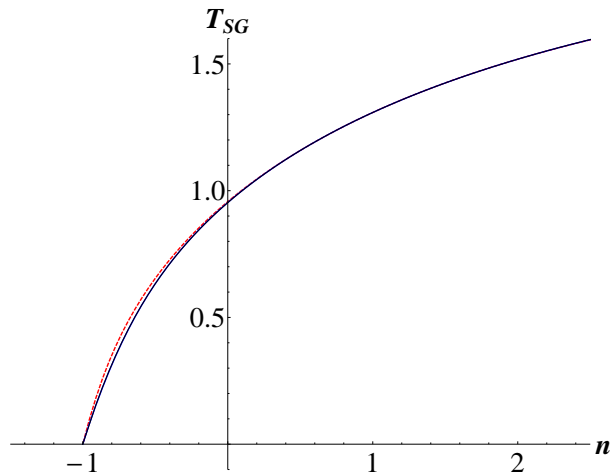}
\end{center}
\caption{{\small Behaviours of $T_{\rm SG}$ as a function of the replica number $n$ derived by $x_0^{(s)} = x_0^{*(s)}$. The dashed curve denotes the results by the naive conjecture, and the solid curve represents those by the improved method for cluster 1. The curve showing the points given by cluster 2 coincides with that by cluster 1 in this scale.}}\label{fig4}
\end{figure}

We show the behaviour of $T_{\rm SG}$ as a function of $n$ derived from $x_0^{(s)} = x_0^{*(s)}$ in figure \ref{fig4} for the naive conjecture and the improved methods. It is clearly observed that the value of $T_{\rm SG}$ depends on the degree of improvement for $n > -1$ but it is fixed to $T_{\rm SG}$ at $n=-1$.

The above arguments are applicable as long as equation (\ref{SG}) can be established.
In this formulation, we used the replica method and the self-duality of lattices.
The present results are acceptable under the validity of the replica method and are applicable to self-dual lattices, the square and several self-dual hierarchical lattices.
Using the star-triangle transformation, we can derive essentially the same relation as equation (\ref{SG}) for the triangular and hexagonal lattices.

We can apply the above formulation to the Gaussian Ising model for $J_0=0$, where $J_0$ denotes the mean of the distribution of randomness because essentially the same improved method works \cite{MCP} and the same relation as equation (\ref{GR}) holds \cite{Georges}.
The same is true for the $q$-state Potts gauge glass and the random Ising model with three-body interactions on the triangular and the Union-Jack lattices.
The multicritical point for these models can be also given by the improved method \cite{NN,MNN,MCP,CC,CC2}.
For instance, the coefficient on the left-hand side of equation (\ref{GR}) is slightly modified and written by the state number $q$ instead of $2$ for the case of the Potts gauge glass.
Also, the duality structure of the four-dimensional random plaquette gauge model is exactly the same as the $\pm J$ Ising model on the square lattice, and therefore the present analysis applies without change \cite{TN,TSN}.
\section{Conclusion and discussions}
We showed the absence of a finite-temperature spin glass transition for several spin glass models with symmetric distribution of randomness in finite dimensions by using the duality and gauge symmetry.
This result is applicable to spin glass models with the Nishimori line, the $\pm J$ Ising model, the Gaussian Ising model, and the $q$-state Potts gauge glass on several self-dual lattices.
We remark that the triangular and hexagonal lattices are included as the self-dual lattices by means of the duality in conjunction with the star-triangle transformation.
The random Ising model with three-body interactions on the triangular and the Union-Jack lattices and and the four-dimensional random plaquette gauge model also have no finite-temperature spin glass transition for the symmetric distribution of randomness.

Many researchers believe the absence of a finite-temperature spin glass transition for the random bond Ising model in two dimensions from numerical investigations.
The present analysis lays an analytical foundation of this expectation.
Although a real-space renormalization group calculation suggests a finite spin glass transition temperature for the triangular lattice \cite{TR}, we believe that the approximation involved there is too crude to be qualitatively reliable.

We showed $T_{\rm SG}=0$ for the symmetric distribution of randomness, considering an arbitrary degree of the improved method.
The conclusion derived in the present analysis is exact under the validity of the improved method in the asymptotic limit of large clusters and the replica method.

Our result is significant in the sense that this is the first analytical and systematic evidence for the conclusion of the absence of a finite-temperature spin glass transition of the random-bond Ising model in two dimensions and related systems with symmetric distribution function of randomness.
We believe that further generalizations to other systems are worth the efforts.
\section*{Acknowledgments}
Fruitful discussions with Tomoyuki Obuchi are gratefully acknowledged.
This work was partially supported by the Ministry of Education,
Science, Sports and Culture, Grant-in-Aid for Young Scientists (B) No.
20740218, and for scientific Research on the Priority Area ``Deepening and
Expansion of Statistical Mechanical Informatics (DEX-SMI)'', and by CREST, JST.

\end{document}